\documentclass[aps,preprint,nofootinbib]{revtex4}%
\usepackage{amsfonts}
\usepackage{amsmath}
\usepackage{amssymb}
\usepackage{graphicx}%
\begin{document}
\title[Quantum ballot]{A quantum secret ballot}
\author{Shahar Dolev}
\email{shahar@dolevim.org}
\author{Itamar Pitowsky}
\affiliation{The Edelstein Center, Levi Building, The Hebrerw University, Givat Ram,
Jerusalem, Israel}
\email{itamarp@vms.huji.ac.il}
\author{Boaz Tamir}
\affiliation{Department of Philosophy of Science, Bar-Ilan University, Ramat-Gan, Israel.}
\email{canjlm@actcom.co.il}

\begin{abstract}
The paper concerns the protection of the secrecy of ballots, so that the
identity of the voters cannot be matched with their vote. To achieve this we
use an entangled quantum state to represent the ballots. Each ballot includes
the identity of the voter, explicitly marked on the "envelope" containing it.
Measuring the content of the envelope yields a random number which reveals no
information about the vote. However, the outcome of the elections can be
unambiguously decided after adding\textit{ }the random numbers from all
envelopes. We consider a few versions of the protocol and their complexity of implementation.

\end{abstract}
\startpage{1}
\maketitle

\section{Introduction}

Assume $n$ parties participate in a vote to decide between a few alternatives.
Each participant chooses his or her preference, designates it on the ballot,
and puts in in the box. There are various ways in which the secrecy of the
vote can be compromised, and we shall be particularly interested in the case
of marked ballots. In this conspiracy the ballot is made to include the
voter's identity, by secretly marking it prior to the vote or during it. To
choose a rather paranoid scenario: Big Brother finds traces of the voter's DNA
on the paper ballot. Or, in the case of an electronic ballot, a string
containing the voter's identity, which has just been varified prior to voting,
is stored together with the vote.

In this paper we use entangled qbits to prevent such schemes. Each ballot may
very well include the identity of the voter, explicitly marked on the
"envelope" containing it. However, this is inconsequential because reading the
contents of the envelope rveals a random number, and no information about the
vote. On the other hand, the outcome of the elections can be unambiguously
decided after \textit{adding }the random numbers from all envelopes.

A few quantum voting protocols have been proposed recently: Singh and Srikanth
\cite{1} suggested to use a quantum version of sealed envelopes; any attempt
to read their content by unauthorised persons can be detected. Vaccaro et.al.
\cite{2} proposed a voting scheme in which the number of votes is coded into
the phases of an entangled state and reading the result involves a complicated
measurement. A protocol more similar to the present one has been proposed by
Hillery et.al. \cite{3}.\ In their protocol the election result is also
encoded into the phases of a quantum state, and its reading involves a
complicated measurement. Our mechanism is different from \cite{3} in various
respects which will be noted below. In particular, the voting result is coded
and read directly from the computation basis states. The protocol can be
implemented as soon as the implementation of the discrete Fourier transform
becomes possible.

We begin with a protocol for a vote to decide between two alternatives.
Although the protocol is valid for any number of voters $n\geq2$, its
implementation may be complicated when thousands of citizens participate in
the elections. To ammend this situation we also propose an alternative
version, whose complexity depends on the number of \textit{ballot boxes}. The
security of the protocol remains intact provided this number is $\geq2$.
Subsequently, the scheme is generalized to include a choice between more than
two alternatives. Finally, the complexity of implementation is calculated.

\section{The Protocol}

Let $m$ be a natural number. Consider an $m$ dimensional space with basis
vectors: $\left\vert 0\right\rangle ,\left\vert 1\right\rangle ,...,\left\vert
m-1\right\rangle $. The $m$-th order discrete Fourier transform is defined to
be%
\begin{equation}
\mathcal{F}_{m}\left\vert j\right\rangle =\frac{1}{\sqrt{m}}\sum_{l=0}%
^{m-1}\exp(\frac{2\pi ijl}{m})\left\vert l\right\rangle ,\qquad j=0,1,...,m-1
\label{1}%
\end{equation}
Subsequently we shall suppress the subscript $m$, and denote the Fourier
transform by $\mathcal{F}$. Let $\Pi$ be the unitary operator which defines
the following cyclic permutation on the basis elements:%

\begin{equation}
\Pi\left\vert 0\right\rangle =\left\vert 1\right\rangle ,\;\Pi\left\vert
1\right\rangle =\left\vert 2\right\rangle ,\;...\;,\Pi\left\vert
m-1\right\rangle =\left\vert 0\right\rangle \label{2}%
\end{equation}
or, in short $\Pi\left\vert j\right\rangle =\left\vert j\oplus1\right\rangle $
where $\oplus$ represents addition $\operatorname{mod}m$.

Suppose that we distribute among $n$ voters the entangled state%

\begin{equation}
\left\vert W\right\rangle =\frac{1}{\sqrt{m}}\sum_{j=0}^{m-1}\left\vert
j\right\rangle \left\vert j\right\rangle ...\left\vert j\right\rangle
\label{3}%
\end{equation}
where each $\left\vert j\right\rangle $ is an $m$-dimensional basis state, and
each product in the sum (\ref{3}) contains $n$ copies. The relation between
$n$ and $m$ will be fixed later. Each voter has to vote either NO, in which
case he applies $\mathcal{F}$ to his bit; or YES, in which case she applies
$\Pi\mathcal{F}$ (that is, $\mathcal{F}$ followed by $\Pi$)\footnote{Hillery
et. al. \cite{3} use the same initial state $\left\vert W\right\rangle $,
apply $\mathcal{F}$ for the YES vote and $I$ (identity) in the NO vote. The
election outcome is then recorded in the phases of a complicated state.}.
Suppose the votes were $a_{1},a_{2},...,a_{n}$, with $a_{k}=0$ in case of a NO
vote by person $k$, and $a_{k}=1$ in case of a YES vote. Put $\Pi^{0}=I$
(identity) and $\Pi^{1}=\Pi$, then after the vote the state is:%

\begin{align}
\left\vert V\right\rangle  &  =(\Pi^{a_{1}}\mathcal{F})\otimes(\Pi^{a_{2}%
}\mathcal{F})\otimes...\otimes(\Pi^{a_{n}}\mathcal{F})\left\vert
W\right\rangle =\label{4}\\
&  =\frac{1}{\sqrt{m}}\sum_{j=0}^{m-1}(\Pi^{a_{1}}\mathcal{F})\left\vert
j\right\rangle \otimes(\Pi^{a_{2}}\mathcal{F})\left\vert j\right\rangle
\otimes...\otimes(\Pi^{a_{n}}\mathcal{F})\left\vert j\right\rangle
=\nonumber\\
&  =\frac{1}{\sqrt{m}}\sum_{j=0}^{m-1}\left(  \Pi^{a_{1}}\frac{1}{\sqrt{m}%
}\sum_{l_{1}=0}^{m-1}\exp(\frac{2\pi ijl_{1}}{m})\left\vert l_{1}\right\rangle
\right)  \otimes...\otimes\left(  \Pi^{a_{n}}\frac{1}{\sqrt{m}}\sum_{l_{n}%
=0}^{m-1}\exp(\frac{2\pi ijl_{n}}{m})\left\vert l_{n}\right\rangle \right)
\nonumber
\end{align}
Performing the tensor product we get:%
\begin{equation}
\left\vert V\right\rangle =\frac{1}{\sqrt{m}}\sum_{j=0}^{m-1}\frac{1}%
{m^{\frac{n}{2}}}\sum_{l_{1},...,l_{n}}\exp\left(  \frac{2\pi ij}{m}%
(l_{1}+...+l_{n})\right)  \left\vert l_{1}\oplus a_{1}\right\rangle
\otimes...\otimes\left\vert l_{n}\oplus a_{n}\right\rangle \label{5}%
\end{equation}
exchanging the order of summation%
\begin{equation}
\left\vert V\right\rangle =\frac{1}{m^{\frac{n+1}{2}}}\sum_{l_{1},...,l_{n}%
}\left(  \sum_{j=0}^{m-1}\exp\left(  \frac{2\pi ij}{m}(l_{1}+...+l_{n}%
)\right)  \right)  \left\vert l_{1}\oplus a_{1}\right\rangle \otimes
...\otimes\left\vert l_{n}\oplus a_{n}\right\rangle \label{6}%
\end{equation}
Unless $l_{1}+...+l_{n}\equiv0(\operatorname{mod}m)$ we have $\sum_{j=0}%
^{m-1}\exp\left(  \frac{2\pi ij}{m}(l_{1}+...+l_{n})\right)  =0$. Hence the
result of the vote is%

\begin{equation}
\left\vert V\right\rangle =\frac{1}{m^{\frac{n-1}{2}}}\sum_{l_{1}%
+...+l_{n}\equiv0(\operatorname{mod}m)}\left\vert l_{1}\oplus a_{1}%
\right\rangle \otimes...\otimes\left\vert l_{n}\oplus a_{n}\right\rangle
\label{7}%
\end{equation}
Now, we measure the basis vectors and add the results $\operatorname{mod}m$.
Since $l_{1}+...+l_{n}\equiv0(\operatorname{mod}m)$ for every component in the
superposition in Eq.(7), we are left with the outcome $a_{1}+...+\nolinebreak
a_{n}(\operatorname{mod}m)$.

\section{Applications}

\textbf{1}. In the simplest case we choose $m>n$, preferably we let $m$ be the
smallest power of two greater than $n$, so we can use qbits. Then, after
adding the measurement results $\operatorname{mod}m$, we simply get
$a_{1}+...+a_{n}$, which is the number of YES votes. The secrecy of the vote
is maintained because every individual "ballot" $\left\vert l_{r}\oplus
a_{r}\right\rangle $ contains the actual vote $a_{r}$ added
$\operatorname{mod}m$ to a random number $l_{r}$ between $0$ and $m-1$. Note
that the ballots are \textit{not} mixed, and it may be public knowledge that
the ballot $\left\vert l_{r}\oplus a_{r}\right\rangle $ comes from voter $r$
(we may even attach an extra probe carrying his or her name). However, this
information is inconsequential, it only reveals the fact that person $r$
participated in the poll.

Actually, we do not have to know in advance how many people will vote, just
choose $n$ to be sufficiently large. Since at the end of election day we know
the exact number of people who participated, we push the NO button as many
times as required to reach $n$. After the measurement we subtract the number
of fictional votes and announce the election results.

There is a classical protocol which is similar to the quantum ballot, but is
nevertheless less secure: A sequence of $n$ random numbers $(l_{1},...,l_{n})$
is generated and their sum $y=l_{1}+...+l_{n}$ stored. When citizen $r$ is
voting $a_{r}$, the electronic voting machine stores only the number
$l_{r}+a_{r}.$ This way the privacy of the vote is protected. At the end of
the day the stored numbers are added, and then $y$ subtracted. This protocol
is secured only to the extent that the values of the random numbers are
protected. In the classical world there is always an interval of time when the
values of the $l_{r}$'s themselves are present in the system. In the quantum
protocol, by contrast, the numbers $l_{r}$ are generated only upon
measurement, and are present only in the compounds $l_{r}\oplus a_{r}$.

Note that an identical result obtains if we change the protocol slightly:
Firstly, we distribute among the voters the state
\begin{equation}
\left\vert U\right\rangle =(\mathcal{F}\otimes\mathcal{F}\otimes
...\otimes\mathcal{F})\left\vert W\right\rangle =\frac{1}{m^{\frac{n-1}{2}}%
}\sum_{l_{1}+...+l_{n}\equiv0(\operatorname{mod}m)}\left\vert l_{1}%
\right\rangle \otimes...\otimes\left\vert l_{n}\right\rangle , \label{8}%
\end{equation}
and secondly, each voter applies $\Pi$ for a YES vote, or $I$ for NO. The
choice between the two versions will depend on the technical detail of implementation.

It goes without saying that even a quantum protocol cannot be secured against
all possible attacks by Big Brother, such as complete rewiring of the voting
machine, or the installation of video cameras in the voting booths.

\textbf{2.} The single element that makes the protocol difficult to execute is
the number of voters $n$. The difficulty is expressed in the structure of the
initial state $\left\vert W\right\rangle $ in Eq.(3), where each component is
a tensor product of $n$ states. If we consider a vote of a small committee
then producing $\left\vert W\right\rangle $ seems feasible; but what if
millions of people vote? Luckily we can simplify the protocol to include this
case. To do this let $N$ stand for the number of \textit{ballot boxes}, and
assume that in the state $\left\vert W\right\rangle $ each component has $N$
copies, one for each box. However, we keep $m$ larger than the total number of
voters\textit{ }$n$. Now, early in the morning on election day, an official
performs $\mathcal{F}$ once for each box, and this is the last time the
Fourier transform is applied to the box. Subsequently, any NO voter applies
$I$ (identity) to the part of the state corresponding to his box, and any YES
voter applies $\Pi$. By repeating the same calculation we get the post
election state%
\begin{equation}
\left\vert V\right\rangle =\frac{1}{m^{\frac{N-1}{2}}}\sum_{l_{1}%
+...+l_{N}\equiv0(\operatorname{mod}m)}\left\vert l_{1}+a_{1}^{\prime
}+...+a_{k_{1}}^{\prime}(\operatorname{mod}m)\right\rangle \otimes
...\otimes\left\vert l_{N}+a_{1}^{\prime\prime}+...+a_{k_{N}}^{\prime\prime
}(\operatorname{mod}m)\right\rangle \label{9}%
\end{equation}
Where $a_{1}^{\prime},...,a_{k_{1}}^{\prime}$, are the votes cast in box $1$,
and so on, to $a_{1}^{\prime\prime},...,a_{k_{N}}^{\prime\prime}$, the votes
in box $N$. Again, since $l_{1}+...+l_{N}\equiv0(\operatorname{mod}m)$, then
measuring the basis states and adding the results $\operatorname{mod}m$ yields
the sum of all YES\ votes from all boxes (recall that we kept $m$ larger than
$n$).

So why not take $N=1$, that is, only one box for all voters? In this case
$\left\vert W\right\rangle =\frac{1}{\sqrt{m}}\sum_{j=0}^{m-1}\left\vert
j\right\rangle $, and $\mathcal{F}\left\vert W\right\rangle =\left\vert
0\right\rangle $, and thus $\left\vert V\right\rangle =\Pi^{a_{1}}\Pi^{a_{2}%
}...\Pi^{a_{n}}\left\vert 0\right\rangle =\left\vert a_{1}+...+a_{n}%
\right\rangle $ is the sum of all YES votes. In other words, using the
protocol with a single ballot box brings us back to a classical voting system
represented by an unentangled quantum state. But already with $N=2$ there is a
random element in the protocol, hiding the number of YES votes that each box contributes.

\textbf{3. }Suppose that there are more than two alternatives, not just YES
and NO, but three candidates to choose from, call them I, II, and III. For $n$
voters we choose $m$ to be bigger than $2n$ and use two copies of $\left\vert
W\right\rangle $, call them $\left\vert W\right\rangle _{1}$ and $\left\vert
W\right\rangle _{2}$. Now, each voter applies the following rule: For
candidate I apply $\mathcal{F}$ to $\left\vert W\right\rangle _{1}$ and
$\mathcal{F}$ to $\left\vert W\right\rangle _{2}$, For candidate II apply
$\Pi\mathcal{F}\left\vert W\right\rangle _{1}$ and $\Pi\mathcal{F}\left\vert
W\right\rangle _{2}$, and for III apply $\Pi^{2}\mathcal{F}\left\vert
W\right\rangle _{1}$ and $\Pi\mathcal{F}\left\vert W\right\rangle _{2}$. Let
$n_{I}$, $n_{II}$ and $n_{III}$ be the numbers of votes cast for the
respective candidates. Applying a measurement to $\left\vert V\right\rangle
_{1}$, the post elections state of $\left\vert W\right\rangle _{1}$, we obtain
the outcome $n_{II}+2n_{III}$. Measuring $\left\vert V\right\rangle _{2}$
yields $n_{II}+n_{III}$. Since we know $n=n_{I}+n_{II}+n_{III}$, we can infer
the election results. Generalizations to a larger number of alternatives is straightforward.

\section{ Complexity of implementation.}

The implementation of the protocol requires three steps:

\textbf{1. The creation of the state }$\left\vert W\right\rangle $. Consider
first the \textit{basis states copier} defined on $\mathbb{C}^{m}%
\otimes...\otimes\mathbb{C}^{m}$ ($n$ copies) and whose effect is, in
particular%
\begin{equation}
\left\vert j\right\rangle \otimes\left\vert 0\right\rangle \otimes
...\otimes\left\vert 0\right\rangle \rightarrow\left\vert j\right\rangle
\otimes\left\vert j\right\rangle \otimes...\otimes\left\vert j\right\rangle
\quad0\leq j\leq m \label{10}%
\end{equation}
To implement this suppose $m=2^{k}$, then the operation $\left\vert
j\right\rangle \otimes\left\vert 0\right\rangle \rightarrow\left\vert
j\right\rangle \otimes\left\vert j\right\rangle $ can be achieved using bit by
bit copying, each bit by the implementation of two CNOT gates \cite{4},
altogether $2k$ gates. Generalizing to $n$ copies we need $O(kn)$ gates,
$k=\log_{2}m$. To create $\left\vert W\right\rangle $, therefore, we apply
this copying mechanism to $(m^{-\frac{1}{2}}\sum_{j=0}^{m-1}\left\vert
j\right\rangle )\otimes\left\vert 0\right\rangle \otimes...\otimes\left\vert
0\right\rangle $, where the state $(m^{-\frac{1}{2}}\sum\left\vert
j\right\rangle )$ is obtained from $\left\vert 0\right\rangle $ by the
application of $k$ Hadamard transforms, one for each bit.

\textbf{2. The application of the Fourier transform }$\mathcal{F}$. By the
central result of Shor \cite{5} $\mathcal{F}$ can be implemented using
$O[(\log m)^{2}]$ gates, and we apply one Fourier transform per copy.
Therefore, the fact that $m$ has to be a large number, larger than the number
of voters $n$, should not pose a big problem. In fact, with $k=25$ binary
digits we can accommodate elections in a mid size country. However, even
smaller scale Fourier transforms suffice to implement an elections protocol
using the following trick: Let $n$ be the number of voters and suppose first
that we know $n$ in advance. Suppose, moreover, that $n=m_{1}m_{2}...m_{s}$,
where $m_{1},m_{2},...,m_{s}$ are coprime. Now, perform the election in
parallel on the $s$ states%

\begin{equation}
\left\vert W\right\rangle _{l}=\frac{1}{\sqrt{m_{l}}}\sum_{j=0}^{m_{l}%
-1}\left\vert j\right\rangle \left\vert j\right\rangle ...\left\vert
j\right\rangle ,\quad1\leq l\leq s \label{11}%
\end{equation}
Where each term in Eq (11) has $n$ copies. Assume that after the measurement
on the $k$ post-election states we get the results $c_{1},c_{2},...,c_{k}$.
The number of YES votes $x$ is satisfying%

\begin{equation}
x=c_{1}(modm_{1}),x=c_{2}(modm_{2}),...,x=c_{s}(modm_{s}), \label{12}%
\end{equation}
and this set of congruences has a unique solution $\operatorname{mod}n$
\cite{6}.

If we do not know $n$ in advance, or if there is no nice decomposition of $n$
to a product of coprimes, we can do as indicated previously: Choose a large
enough $n$ to be on the safe side, and make sure it has a comfortable
decomposition. After election day is over push the NO button as many time as
needed to bring the number of votes to $n$, and subsequently subtract the
fictional votes before the result is announced.

\textbf{3. Application of} $\Pi$: Is just an implementation of an algorithm
that performs $j\rightarrow j+1$ $\operatorname{mod}m$, which takes $O(\log
m)$ steps per copy.

Altogether, the complexity of the protocol is $O[n(\log_{2}m)^{2}]$ where $n$
is the number of voters (or in another scheme, the number of ballot boxes) and
$m$ is the least power of two greater than the number of voters.

\end{document}